\documentclass{PoS}
\let\ifpdf\relax

\title{Photometry of the Stellar Tidal Stream in the Halo of Messier 63
\footnote{This contribution includes data taken at The McDonald Observatory of The University of Texas at Austin.}
\footnote{Following its presentation at this symposium, this research was published unabridged \cite{chonis2011}.}}

\ShortTitle{Photometry of the Stellar Tidal Stream in the Halo of Messier 63}

\author{
\speaker{Taylor S. Chonis}$^{,a}$, David Mart\'{i}nez-Delgado$^{b}$, R. Jay Gabany$^{c}$, Steven R. Majewski$^{d}$, Gary J. Hill$^{e}$, Ignacio Trujillo$^{f,g}$\\
\llap{$^{a}$}Department of Astronomy, University of Texas at Austin\\  
\llap{$^{b}$}Max-Planck Institut f\"{u}r Astronomie\\
\llap{$^{c}$}BlackBird Observatory\\
\llap{$^{d}$}Department of Astronomy, University of Virginia\\
\llap{$^{e}$}McDonald Observatory, University of Texas at Austin\\
\llap{$^{f}$}Instituto de Astrof\'{i}sica de Canarias\\
\llap{$^{g}$}Departmento de Astrof\'{i}sica, Universidad de La Laguna\\
E-mail: \email{tschonis@astro.as.utexas.edu}, \email{delgado@mpia-hd.mpg.de}, \email{rj2010@cosmotography.com}, \email{srm4n@virginia.edu}, \email{hill@astro.as.utexas.edu}, \email{itc@iac.es} 
}

\abstract{We present surface photometry of a giant, low surface brightness stellar arc in the halo of the nearby spiral galaxy M63 (NGC 5055) that is consistent with being a part of a stellar stream resulting from the disruption of a dwarf satellite galaxy. Using the stream's ``great-circle'' morphology and its photometric properties, we estimate that the stream originates from the accretion of a $\sim10^{8}$ $M_{\odot}$ satellite in the last few Gyr. The $B-R$ color of the stream's stars is consistent with Local Group dwarfs and is also similar to the outer regions of M63's disk and stellar halo within our measurement uncertainties. Additionally, we identify several other low surface brightness features that may be related to the galaxy's complex spiral structure or may be tidal debris associated with the disruption of the galaxy's outer stellar disk as a result of the accretion event. Using our deep, panoramic optical view of M63 with additional existing multiwavelength data, we describe the possible effects of such an accretion event in the larger picture of the parent galaxy.}

\FullConference{Frank N. Bash Symposium 2011: New Horizons in Astronomy\\
                 October 9-11, 2011 \\
                 Austin, Texas, USA}

\def\arcsec{^{\prime\prime}}
\def\arcmin{^\prime}

\begin{document}

\section{Introduction}
In the $\Lambda$CDM paradigm, the merging of dark matter halos drives the evolution of galaxies. In the present epoch, minor mergers, which often alter stellar disks and help build stellar halos from the inside-out, are expected to be common. The signatures of these interactions (such as heated disks, warps, and stellar tidal debris) should  be visible for several Gyr. Recent cosmological simulations predict that most large spiral galaxies should show signs of minor interactions if observed to sufficient depth (e.g., $\mu_{V} \sim 30$ mag arcsec$^{-2}$) \cite{johnston2008,cooper2010}. Several Local Volume spirals display signatures of recent minor mergers and show striking qualitative agreement with the cosmological models \cite{martinezDelgado2010}. Here, we present deep follow-up surface photometry of the outer regions of M63, a galaxy that displays clear evidence of a recent minor merger \cite{chonis2011}. We aim to determine the basic properties of what remains of the progenitor satellite galaxy and study the effects of the accretion event on the parent system. M63 is an isolated, flocculent SA(rs)bc spiral. It is 7.2 Mpc distant, has a stellar mass of $\sim8\times10^{10}$ $M_{\odot}$ at $<40$ kpc, displays a pronounced HI warp, and hosts a Type 1 extended UV (XUV) disk \cite{battaglia2006, thilker2007}. 

\section{Observations}
We obtained 6.5 (5.33) hrs of total exposure through $B$ ($R$) filters with the McDonald Observatory 0.8 m telescope and PFC imager ($46.2\arcmin\times42.6\arcmin$ FOV at 1.354$\arcsec$ px$^{-1}$). A detailed description of the data analysis can be found in \cite{chonis2011}. Standard data reduction methods were employed. We utilize a row-by-row and column-by-column background fitting routine with low-order polynomials to adaptively fit and subtract the background. ``Tertiary'' SDSS standards in the target field were used for photometric calibration. We correct for Galactic extinction, which is uniform over the FOV to $<0.01$ mag. Through an extensive estimation of uncertainties, we find that random noise is insignificant for a typical measurement while the background subtraction is the limiting systematic uncertainty. A measurement of a diffuse feature with $\mu_{R} = 26.0$ has an uncertainty of $\pm0.1$ mag arcsec$^{-2}$. Figure \ref{fig1}$a$ shows the final $R$-band image. Also shown are our final data products: an $R$-band surface brightness map (\ref{fig1}$b$) and a $B-R$ color index map (\ref{fig1}$c$) for features with $23 \lesssim \mu_{R} \lesssim 27$.

\section{The Stellar Stream}
\begin{figure}
\centerline{\includegraphics[width=\textwidth]{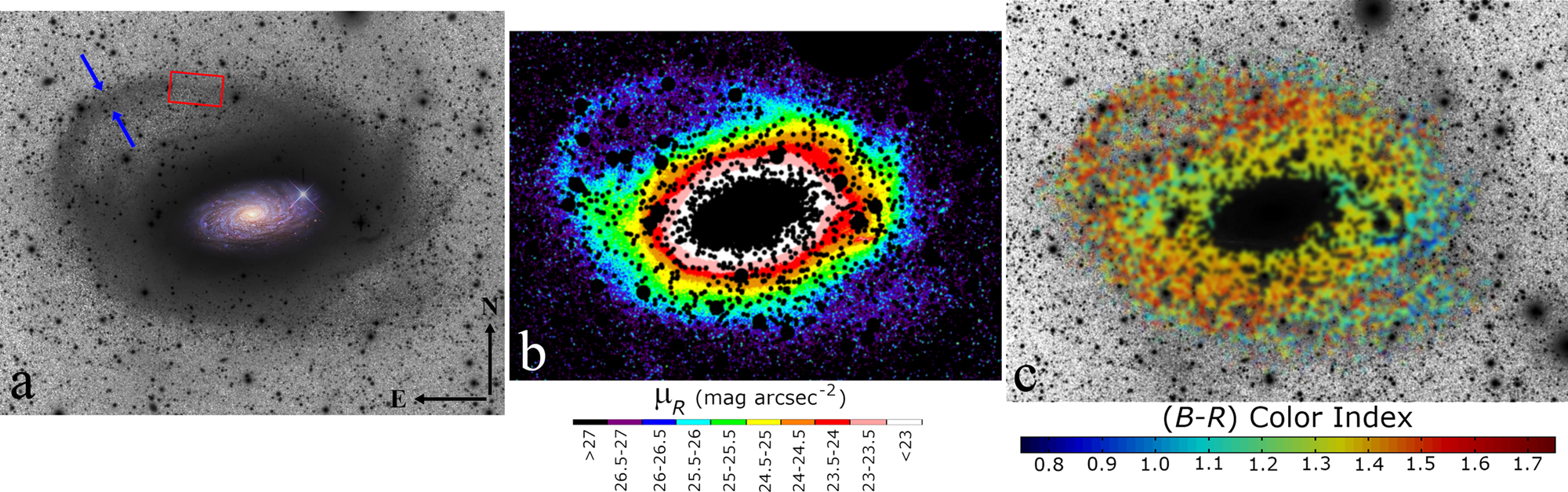}}
\caption{Optical data: \textbf{Panel \emph{a}} - The final $R$-band image of M63. The directional arrows in the lower-right corner are 6$\arcmin$ (12.6 kpc) long. The blue arrows indicate the location along the stream where $w$ is measured. The red box indicates the location of a dim break in the stream that is $\sim 1$ mag arcsec$^{-2}$ dimmer than the stream's average $\mu_{R}$. A tri-color overlay of the M63 stellar disk is also shown \cite{chonis2011}. \textbf{Panel \emph{b}} - $R$-band surface brightness contour map derived from panel $a$ for $23 \lesssim \mu_{R} \lesssim 27$. \textbf{Panel \emph{c}} - $B-R$ color index map overlayed on the $R$-band image. Typical errors in $B-R$ are $\pm0.2$ mag for a feature with $\mu_{R} \approx 26$.}
\label{fig1}
\end{figure}

The most striking low surface brightness feature we observe is the coherent arc structure that reaches $R\approx29$ kpc (in projection) from the center of M63 (see Figure \ref{fig1}). The arc is reminiscent of a ``great-circle'' stellar tidal stream that would result from a tidally disrupted satellite galaxy \cite{johnston2008}. An ellipse fit to the light distribution displays a $14.8^{\circ}$ tilt with respect to the inner disk isophotes. The FWHM of the stream's light distribution, $w$, is 3.3 kpc (in projection). The stream has an average $\mu_{R} = 26.1 \pm 0.1$ and $B-R = 1.5\pm0.2$ within the FWHM, the latter of which is consistent with an average S0 galaxy \cite{barway2005} and redder Local Group dwarfs \cite{mateo1998}. It is also consistent with the $B-R$ color of the outer disk of M63 at the $\mu_{R} = 23.5$ isophote within the 1$\sigma$ measurement uncertainty. In Figure \ref{fig2}, it can be seen that existing multiwavelength data shows that there is little to no HI gas ($< 0.10$ $M_{\odot}$ pc$^{-2}$) or UV flux associated with the stellar stream. Along with the red $B-R$ color measured in this work, these data suggest that the stream is composed of an old stellar population.

In addition to the coherent arc, we detect other low surface brightness features (see Figure \ref{fig1}). Some of these correspond to bright knots of UV flux associated with recent star formation in the XUV disk (see Figure \ref{fig2}$b$; all such features have $B-R < 1.1$). Several other faint features have red $B-R$ colors that are similar to the outer disk isophotes and stellar stream and appear asymmetric in an azimuthal manner about the inner stellar disk. 

\section{Discussion}
$\bullet$ \emph{Origin of the Faint Light Features} - We have presented data that make a strong case for the faint features around M63 being the result of an ongoing minor accretion event involving a satellite galaxy. The stream's morphology resembles the ``great-circle'' stellar streams in $\Lambda$CDM simulations \cite{johnston2008, cooper2010}. ``Great-circle'' streams that are still-bound result from recent accretion events ($<6$ Gyr) where the progenitor is on an orbit with only mild eccentricity \cite{johnston2008}. The stream's $\mu_{R}$ is among the brightest of those in the simulations \cite{johnston2008}, which suggests a very recent disruption of the progenitor satellite.

$\bullet$ \emph{Fate of the Progenitor Dwarf} - Our data do not provide the current position or fate of a remaining bound core of the progenitor dwarf galaxy. It could be hidden behind or superposed on M63's stellar disk. While there are several faint features with redder $B-R$ colors that could be viable candidates, it is more likely that any remaining bound core is indiscernible in the stream since the core surface brightness decreases monotonically after the initial disruption. 

$\bullet$ \emph{Estimation of the Stream's Basic Dynamical Properties} - We estimate the current stream mass, $m$, and the time since the initial disruption, $t$, from the stream's morphological characteristics using the analytic framework of \cite{johnston2001}. Assuming a near circular orbit and using $R$ and $w$ measured from our data along with $v_{circ}(R) \approx 180$ km s$^{-1}$ \cite{battaglia2006}, we find that $m \approx 3.5\times10^{8}$ $M_{\odot}$. This is similar to several Local Group dSphs \cite{mateo1998} and is consistent with the prominent satellites from $1<z<7$ that assemble stellar halos in the $\Lambda$CDM simulations \cite{cooper2010}. The time since disruption depends linearly on $\Psi$, the stream's angular length; however, $\Psi$ is not conclusive from our data. We thus use the parameterization $\Psi = 2\pi\eta$ (where $\eta$ is the number of wraps the stream makes around M63) so that $t \approx 1.8\eta$ Gyr. As an alternate estimation of $m$, we measure the stream's surface luminosity density $\Sigma_{R}$ (in units of $L_{R \odot}$ pc$^{-2}$) and estimate a mass-to-light ratio, $\frac{m}{L_{R}}$, from the stream's $B-R$ color \cite{bell2001}. Using the results of the ellipse fit to the stream's light distribution, a sky projected area, $A$, is estimated (in units of pc$^{2}$). Using the $\eta$ parameterization within $A$, we find that $m = \Sigma_{R} A (\frac{m}{L_{R}}) \approx (4 \pm 2)\eta \times 10^{8}$ $M_{\odot}$. Our data only clearly show a single coherent arc (i.e.,  $\eta \gtrsim 0.5$). The dim break in the otherwise coherent stellar stream (which may indicate the location of the ends of the leading and trailing tidal arms; see Figure \ref{fig1}$a$) as well as the color distribution about M63 (which is redder on the eastern half of the galaxy where possibly the stream is above M63's disk plane; see Figure \ref{fig1}$c$) possibly suggests only a single loop. If this situation is the case,  $\eta \approx 1$, and the two estimations of $m$ are subsequently in agreement. From our data, there is no evidence supporting additional stream wraps.

\begin{figure}
\centerline{\includegraphics[width=\textwidth]{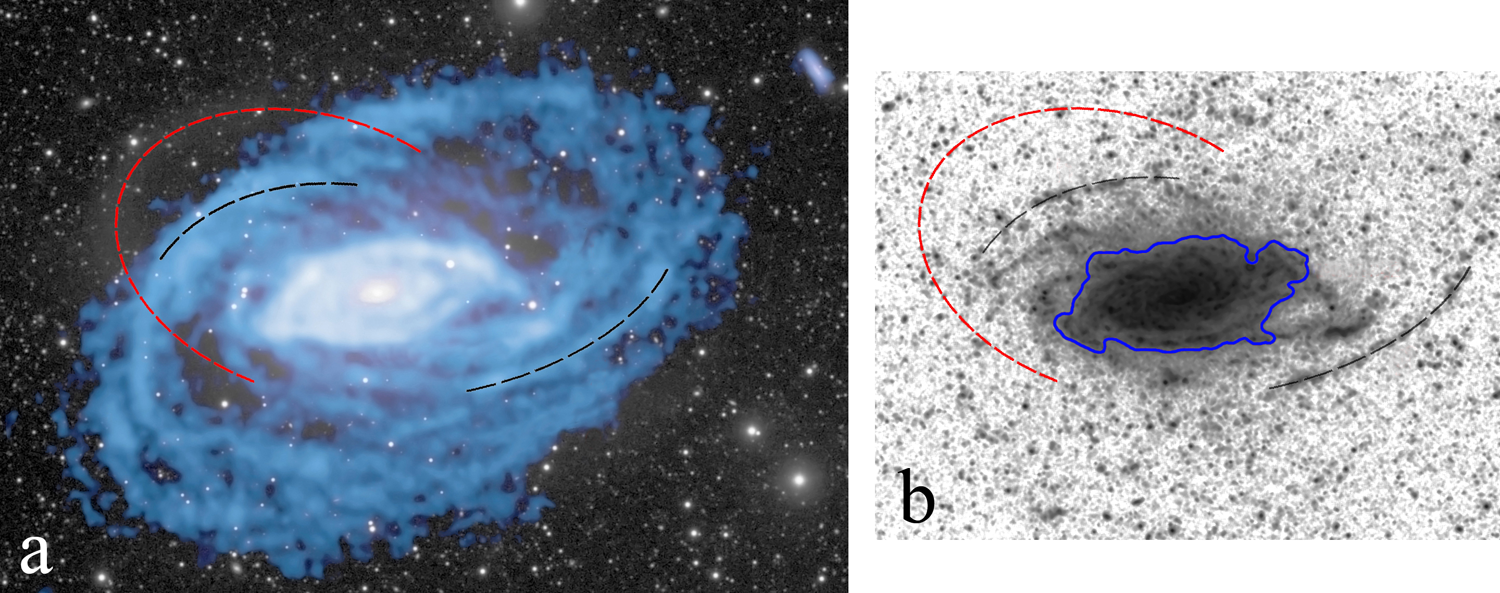}}
\caption{Multiwavelength data: \textbf{Panel \emph{a}} - Comparison of the optical image (grayscale) with the distribution of HI gas (blue) at 67$\arcsec$ resolution \cite{battaglia2006}. The detection limit is 0.10 $M_{\odot}$ pc$^{-2}$. \textbf{Panel \emph{b}} - Archival \emph{GALEX} FUV+NUV image. The blue contour represents the $\mu_{FUV} = 27.25$ AB mag arcsec$^{-2}$ isophote \cite{thilker2007}. In both panels, the red dashed curve represents a segment of the stream ellipse fit. The black curve segments outline examples of features in the XUV disk that can be associated with HI spiral structure.}
\label{fig2}
\end{figure}

$\bullet$ \emph{Effect on the Parent System} - The data hints at the inside-out formation of the stellar halo. The $B-R$ color similarity between the outer disk and the stream is consistent with recent $\Lambda$CDM simulations that show that a stellar halo after a typical minor merger can consist of a mixture of accreted stream stars \emph{and} ejected disk stars \cite{purcell2010}. This could homogenize the integrated color of the stars in the two structures. Several of the faint, redder, and azimuthally asymmetric features we detect could be stars ejected from the disk. These stars may eventually settle into a thick disk component. The accretion event could also be responsible for the large HI warp and the extended spiral structure in the HI disk \cite{battaglia2006} (see Figure \ref{fig2}$a$). The knots of star formation at large radii within these spiral arms that make up the XUV disk could also be caused by instabilities due to the ongoing accretion event \cite{thilker2007} (see Figure \ref{fig2}$b$).\\

This research is a useful step in studying these elusive, yet important pieces of the galactic evolutionary and cosmological puzzles. Similar observations in statistically significant samples will provide an important test of the cosmological simulations that are based on the $\Lambda$CDM paradigm \cite{martinezDelgado2010}. However, this work can only bring limited results, as sky projected stream morphologies and broad-band colors are not sufficient to fully constrain the dynamical and compositional properties of the stellar debris left over from the hierarchical evolutionary process. Future advancement in observational techniques and instrumentation will be needed to reasonably probe these extremely faint structures spectroscopically. This will better allow the use of stellar streams around external galaxies as probes of dark matter halos, as tools for studying the effect of minor mergers on galaxy evolution, and as a means of placing more stringent constraints on future cosmological simulations. Nevertheless, deep imaging of relatively isolated, Local Volume spiral galaxies has shown that their stellar halos still contain the relics from their hierarchical, inside-out formation. This presents a unique opportunity to be witnesses of one of the latest stages of galaxy evolution.

\acknowledgments
This work is partially supported by the Texas Norman Hackerman Advanced Research Program under grant 003658-0295-2007. We thank The McDonald Observatory and its staff for supporting the photometric observations. We acknowledge Ray Gralak for his contribution to the project, but whose data could not be included in this abbreviated proceeding.


\end{document}